\documentclass[aps,showpacs, nofootinbib]{revtex4}

\usepackage{graphicx}

\newcommand{\be}{\begin{equation}}
\newcommand{\ee}{\end{equation}}
\newcommand{\ben}{\begin{eqnarray}}
\newcommand{\een}{\end{eqnarray}}

\newcommand{\la}{{\lambda}}

\newcommand{\om}{{\omega}}

\newcommand{\cA}{{\cal A}}
\newcommand{\cB}{{\cal B}}

\newcommand{\p}{\partial}
\newcommand{\na}{\nabla}

\newcommand{\tom}{{\tilde \omega}}
\newcommand{\tOm}{{\tilde \Omega}}

\newcommand{\ga}{\gamma}

\newcommand{\Dsl}{{\slash \negthinspace \negthinspace \negthinspace \negthinspace  D}}


\begin{document}

\title{Massive Fermion Emission from Higher Dimensional Black Holes} 

\author{Marek Rogatko and Agnieszka Szyp\l owska}
\affiliation{Institute of Physics \protect \\
Maria Curie-Sklodowska University \protect \\
20-031 Lublin, pl.~Marii Curie-Sklodowskiej 1, Poland \protect \\
rogat@kft.umcs.lublin.pl\protect \\
marek.rogatko@poczta.umcs.lublin.pl}

\date{\today}

\begin{abstract}
We study the effect of extra dimensions on the process
of massive Dirac fermion emission in the spacetime of $(4 + n)$-dimensional
black hole, by 
examining the 
Dirac operator in arbitrary
spacetime dimension. We comment on both bulk and brane emission and find
absorption cross section and luminosity of Hawking radiation
in the low-energy approximation.
\end{abstract}

\maketitle

\section{Introduction}
Scientists have devoted many years to the ongoing
quest to unify the forces in Nature.
Higher-dimensional theories 
provide a promising framework for the unification of gravitation
with other fundamental forces. In this context 
braneworld models \cite{ark98,ran99} with large extra dimensions point us the way out of 
the long standing hierarchy problem by lowering the
fundamental scale of gravity down to order of TeV. It has been also argued that
mini black holes might be created through high-energy particle collisions at
TeV-energy scales. These objects are expected to evaporate through Hawking radiation both in
the bulk as well as on the brane. Mini black holes created in high-energy collisions will undergo
a number of phases, i.e. {\it balding phase} when black hole will emit mainly gravitational radiation,
{\it spin-down phase} during which black hole will loose its angular momentum through emission of Hawking radiation
and {\it Schwarzschild phase} in which black hole will lose its actual mass by Hawking radiation.
Finally in {\it Planck phase} quantum gravity theory is needed to study its behaviour.\\
The TeV scale gravity opens up the possibility of producing black holes
and observing their decay product. The aforementioned
range of energy will be soon achieved by CERN Large Hadron Collider. One hopes that it proves or 
merely restricts the parameter range (e.g., number or {\it size} of extra dimensions) of higher
dimensional theories.
\par
Studies of particle emission from multidimensional black hole have their own long history.
Namely, in Ref.\cite{kan03a} massless scalar emission were studied in the spacetime
of $(4 + n)$-dimensional Schwarzschild black hole while
the case of massless spinor and gauge particles was treated in \cite{kan03b}. 
Then, radiation emitted from higher-dimensional black holes were 
considered both analytically and numerically
(see, e.g., Refs.\cite{par06}-\cite{car06} for a non-exhaustive sampling of this widely treated subject).
On the other hand,
graviton emission in the bulk from a higher dimensional Schwarzschild
black hole was elaborated in Refs.\cite{cre06}, where it was established that the low-energy emission rate
decreases with the number of extra-dimensions as was previously found for the case of bulk
massless scalar field.
\par
The complexity of the aforementioned problem in the background of a rotating $(4 + n)$-dimensional black hole
was revealed in Refs.\cite{cre07}.
The argument of the effect of mass on the emission spectrum in four-dimensions was quoted 
in \cite{gra08}
(see also
\cite{fro03} where the particle and light motion in the vicinity of five-dimensional
rotating black hole was investigated).
In Ref.\cite{cas08} the scalar massless Hawking emission into the bulk by a higher dimensional rotating black hole 
with only one non-zero angular momentum component in a plane parallel to the brane
was investigated. 
It was found that 
the bulk emission remained significantly smaller than the brane one. It turned out also that the angular momentum loss
rate had smaller value in the bulk than on a brane.
Fermion emission in $(4 + n)$-dimensional
rotating background obtained by projecting the higher dimensional black hole line element on the brane was studied in \cite{cas07}
where was revealed that any increase of dimensionality or angular momentum parameter of the black hole significantly 
enhanced all emission rates. The same conclusions were drawn in Ref.\cite{ida06}.
It was shown in \cite{cho08},
that for spacetime of dimensionality greater than five bulk fermion emission dominated
brane localized one, contrary to the conjecture that black holes radiated mainly on the brane.
Recently, greybody factor of Schwarzschild black hole localized on a three-brane of finite tension,
both in the low and high frequencies regimes were studied \cite{usa09}.
\par
The absorption probabilities of massless scalar field were also studied in more exotic 
backgrounds, e.g., in the spacetime of
a rotating G\"odel
black hole in minimal five-dimensional gauged supergravity \cite{che08} as well as in the spacetime of rotating
Kaluza-Klein black hole with squashed horizon \cite{che08b}. The Hawking radiation of $(4 + n)$-dimensional
Schwarzschild black hole imbedded in de Sitter spacetime was investigated in Ref.\cite{wu08}.
The problem
of massless scalar emission in the background of $n$-dimensional static black hole
surrounded by quintessence was elaborated in Ref.\cite{che08c}.

\par 
On the other hand, the emission spectrum of minimally coupled particles with spin which were Hawking radiated
from four-dimensional black hole in string theory was studied in \cite{cve98}, while
the low-energy cross section for minimally coupled massless fermions was provided in Ref.\cite{das97}.
The low-energy absorption cross section for massive fermions in the Schwarzschild background defined on the bulk
was investigated in \cite{jun04} using Dirac equation in the traditional form.
\par
Our present work will be devoted to 
$(4 + n)$-dimensional black holes evaporating massive Dirac degrees of freedom.
In our research we use the much easier method of solving massive Dirac equation than the traditional one.
We shall restrict our attention to the static spherically symmetric case.
It turned out that the treatment of fermions in spherically symmetric background
may be simplified by using a few basic properties of the Dirac operator.
These features enable us to find the second order differential equation which 
will be crucial in further investigations.
In Sec.II we recall the basic features of Dirac operator which enable us to treat
the aforementioned problem of massive Dirac fermions in arbitrary spacetime
dimension.
In Sec.III we shall focus on the low-energy regime and solve analytically 
Dirac massive fermion field equations by means of 
the {\it matching technique} combining the far field and near event horizon solutions.
We find the analytical expression for the absorption probability and luminosity of
Hawking radiation. We shall comment both on {\it bulk} and 
{\it brane} fermion emissions. Our analytical considerations are supplemented by 
plots expressing the dependences of the found quantities on various parameters of the considered spacetimes.
Finally, in Sec.IV we shall state our conclusions.

\section{Properties of the Massive Dirac Equation}
In this section we recall the properties of Dirac operator (for the readers convenience) which enable us to
simplify studies of the Hawking massive fermion radiation in higher dimensional black hole backgrounds.
It happened that the treatment of fermions in spherical spacetime may be greatly simplified
due to the few properties of the Dirac equation of motion. This attitude
was widely used in studies of the late-time behaviour of massive Dirac hair 
in the spacetimes of various black holes (see, e.g., \cite{gib08a}-\cite{mod08}). One should remark that our considerations are conducted
in arbitrary spacetime dimension.
\par
The massive Dirac equation in a curved background may be written in the form as
(for the convention we used see Refs.\cite{gib08a,zum62})
\be \bigg( \ga^{\mu} \na_{\mu} - m \bigg) \psi = 0, \ee
where $\na_{\mu}$ is the covariant derivative $\na_{\mu} = \p_{\mu} +
{1\over 4} \omega_{\mu}^{ab} \ga_{a} \ga_{b}$, $\mu$ and $a$ are
tangent and spacetime indices.  They are related by $e_{\mu}^{a}$ with a
basis of orthonormal one-forms.  The quantity $\omega_{\mu}^{ab}
\equiv \omega^{ab}$ is associated with connection one-forms satisfying
$de^{a} + \omega_{b}{}{}^{a} \wedge e^{b} = 0$, while the gamma
matrices fulfill the relation $\{ \ga^{a}, \ga^{b} \} = 2 \eta^{ab}$.
Now, we shall recall some basic properties of the Dirac operator $\Dsl
= \gamma ^\mu \nabla _\mu $ on an $n$-dimensional manifold.
\par
In what follows we assume that the metric of the
underlying spacetime may be rewritten as a product of the form
\be g_{\mu \nu} d x ^\mu d x ^\nu = g_{ab} (x) dx^a dx ^b + g_{mn}(y)
dy^m dy ^n.  \ee
The above metric decomposition will be subject to the direct sum of
the Dirac operator, namely one obtains
\be \Dsl = \Dsl_x + \Dsl_y.  \ee
By virtue of a Weyl conformal rescaling defined by the following:
\be 
g_{\mu \nu} = \tOm ^2 {\tilde g} _{\mu \nu}, 
\ee
where $\tOm$ is a conformal factor,
the above considerations consequently provide that one gets
\be 
\Dsl \psi =\tOm ^{- {1 \over 2} (n+1)} { \tilde {\Dsl} } {\tilde
\psi}\,, \qquad \psi = \tOm ^ {- {1 \over 2} (n-1) }\tilde \psi.
\ee
Having in mind that spherically symmetric form of the line element
provides also conformal flatness for a static metric, one obtains
\be ds^2 = - A^2 dt^2 + B^2 dr ^2 + C^2 d \Omega ^2 _{n+2} \,,  \ee
where $A=A(r)$, $B=B(r)$, $C=C(r)$ are functions only of the radial
variable $r$, and the {\it transverse} metric $d \Omega ^2 _{n+2} $ is
independent on $t$ and on $r$-coordinates.

\noindent
Suppose then, that $\Psi$ is a spinor eigenfunction on the
$(n+2)$-dimensional {\it transverse} manifold $\Omega$.  It leads to
the relation
\be 
\Dsl_\Omega \Psi = \lambda \Psi.  
\ee
Using the properties given above one may also assume that the
following is satisfied:
\be \Dsl \psi = m \psi.  \ee
It enables us to set the form of the spinor $\psi$, i.e.,
\be \psi = {1 \over A^{1 \over 2} } { 1 \over C^{(n-2) \over 2 }} \chi
\otimes \Psi.  \ee
Next, the explicit calculations reveal 
\be (\gamma ^0 \partial _t + \gamma ^1 \partial _y) \chi = A (m-
    {\lambda \over C} ) \chi \,, \ee
where we have introduced the {\it radial optical distance} (i.e., the
Regge-Wheeler radial coordinate) $dy = {B / A} dr$ and $\gamma^0,
\gamma^1$ satisfy the Clifford algebra in two spacetime
dimensions.\par

\noindent
We remark that an identical result may be obtained if a Yang-Mills
gauge field $A_{\mu}$ is present on the {\it transverse} manifold
$\Sigma$.  The only difference is that
\be \Dsl_{\Omega, A_{\mu}} \Psi = \lambda \Psi \,, \ee
where $ \Dsl_{\Sigma, A_{\mu}} $ is the Dirac operator twisted by the
the connection $A_{\mu}$.  
Thus, in the picture under consideration,
assuming that $\psi \propto e^{-i\omega t}$
one achieves the second order equation of the form as follows:
\be {d^2 \chi \over d y^2 }+ \omega ^2 \chi = A^2 \bigg(m-{\lambda \over C}
\bigg)^2 \chi.
\label{second}
\ee

\section{Greybody Factor in the Low-energy Regime}
In this section we shall concentrate our attention on finding the absorption probability in the low-energy regime.
Greybody factors enable us to study the near horizon structure
of black holes. This is of a great 
importance from the experimental point of view due to the fact that they
modify the spectrum in the region of particle production. In general the
spectrum of emitted particles depends on
various factors such as spin of the particles, whether the particle is localized on brane 
or can propagate in the bulk. The greybody factor can be computed by finding the absorption
cross section for the type of particle, in question, incident on the adequate
black hole. This can be done in such a way because of the fact that Hawking's formula for the emission
rate for an outgoing particle at energy $\om$ equals the absorption cross section
for the same type of particle incoming at energy $\om$. Moreover,
outgoing transmission and ingoing absorption coefficients are equal. 
Therefore equilibrium still takes place if the black hole is located in a heat bath.
\par
In what follows we shall investigate massive Dirac fermion emission in the background of
$(4 + n)$-dimensional black hole which
line element of such a black hole is subject to the relation
\be
ds^2 = - f(r)dt^2 + {dr^2 \over f(r)} + r^2 d\Omega^2_{n + 2},
\ee
where $f = 1 - \bigg( r_{0} / r \bigg)^{n + 1}$, $r_{0}$ is the radius of the black hole event horizon,
while $d\Omega^2_{n + 2}$ is a line element
on $S^{n + 2}$ sphere provided by the relation
\be
d\Omega^2_{n + 2} = d\theta^2 + \sum_{i = 2}^{n + 2} \prod_{j =1 }^{i -1} \sin^2 \phi_{j} d\phi_{i}^2.
\ee

Relation (\ref{second}) will constitute the 
defining
equation for massive Dirac fermion fields. 
By virtue of {\it approximation technique} 
we want to achieve the analytical solution of the underlying equation.
Namely, we solve the equation for $\chi$ in the near horizon region
and then in the far-field limit. Our next task is to match them smoothly.
\par
Let us begin with the near horizon limit. After changing of variables in Eq.(\ref{second})
we arrive at the following:
\be
f (1 - f) {d^2 \chi \over df^2} + \bigg[ 1 - (1 + \xi)f \bigg]~{d\chi \over df}
+ \bigg[ 
{\om^2 r^2 \over (n + 1)^2~ f (1 - f)} - {r^2 (m-{\lambda \over C})^2 \over (n + 1)^2~(1 - f)}
\bigg]~\chi = 0,
\ee
where $\xi = {n + 2 \over n + 1}$.\\
Then, one redefines $\chi(f) = f^{\alpha} (1 - f)^{\beta} F(f)$ and remove
singularities at $f = 0$ and $f = 1$. The above redefinition make it possible to
transform this equation to the hypergeometric one of the form as follows:
\be
f~(1 - f)~{d^2 F \over df^2} + [c - (1 + a + b)f]~{dF \over df} - ab F = 0.
\label{a1}
\ee
It can be checked that the hypergeometric equation parameters satisfy
$a = \alpha + \beta + \xi$,~$b = \alpha + \beta$ and $c = 1 + 2 \alpha$, while $\alpha$ and $\beta$
are given by
\ben
\alpha_{\pm} &=& \pm {i \om r_{0} \over n + 1}, \\ \nonumber
\beta_{\pm} &=& - {(\xi -1) \over 2} \pm {\sqrt{ \Delta} \over 2},
\een
where by $\Delta$ we denoted the following:
\be
\Delta = {1 \over (n + 1)^2} - 4 {( \om^2 - (m-{\lambda \over C})^2 )~{r_{0}}^2 \over (n + 1)^2}.
\ee
From this stage on, we shall suppose
for simplicity that $C(r) = r$. Further on,
having in mind the criterion for the hyperbolic function to be convergent, i.e.,
$Re( c - a -b) > 0$, one has to select $\beta = \beta_{-}$. Just the general solution
of Eq.(\ref{a1}) may be written 
in the form
\be
\chi_{NH}(f) = A_{-}~f^{\alpha}~(1 - f)^{\beta}~F(\alpha + \beta + \xi, \alpha + \beta, 1 + 2 \alpha; f)
+ A_{+}~f^{- \alpha}~(1 - f)^{\beta}~F(\beta + \xi - \alpha, \beta -\alpha, 1 - 2 \alpha; f),
\ee
where $A_{\pm}$ are arbitrary constants.
Because of the fact that no outgoing mode exists near
the event horizon of the considered black hole, we take $\alpha = \alpha_{-}$ and
put $A_{+}$ equal to zero. This leads to the following solution of equations of motion:
\be    
\chi_{NH}(f) = A_{-}~f^{\alpha}~(1 - f)^{\beta}~F(\alpha + \beta + \xi, \alpha + \beta, 1 + 2 \alpha; f).
\label{rr1}
\ee
Our next task is to match smoothly the near horizon solution $\chi_{NH}$ with the far field
one in the intermediate zone. To do this, first we change the expression of the hypergeometric
function near horizon zone from $f$ to $(1- f)$ by the standard relation (see \cite{abr66} relation $15.3.6$).
\par
Having in mind the limit $r \rightarrow \infty$ or adequately $f \rightarrow 1$ and assuming 
the low-energy limit $\tom \ll 1$, we are finally left with
\be
\chi_{FF} = B_{+}~\sqrt{r}~J_{{1 \over 2}\sqrt{1 + 4 \la^2}}(\tom r)
+ B_{-}~\sqrt{r}~Y_{{1 \over 2}\sqrt{1 + 4 \la^2}}(\tom r),
\ee
where $Y_{\nu}$ and $J_{\nu}$ are Bessel functions of second and first order, respectively.\\
On expanding in the limit $\tom r \rightarrow 0$ we conclude that
\be
\chi_{FF}(\tom r \rightarrow 0) = B_{+}~
{\big( {1 \over 2} \tom \big)^{{1 \over 2}\sqrt{1 + 4 \la^2}}
\over
\Gamma( 1 + {1 \over 2}\sqrt{1 + 4 \la^2})}~ \big( r \big)^{{1 \over 2} + {1 \over 2}\sqrt{1 + 4 \la^2}}
- {B_{-} \over \pi}~\Gamma( {1 \over 2}\sqrt{1 + 4 \la^2})~
\big( {1 \over 2} \tom \big)^{- {1 \over 2}\sqrt{1 + 4 \la^2}}~
\big( r \big)^{{1 \over 2} - {1 \over 2}\sqrt{1 + 4 \la^2}}.
\ee
Consequently, taking the near-horizon limit for $\chi_{NH}$, one gets
\ben
\chi_{NH}(f \rightarrow 1) &=& A_{-} \bigg[
\big( r_{0} \big)^{- {1 \over 2} - {1 \over 2}\sqrt{1 + 4 \la^2}}~
\big( r \big)^{{1 \over 2} + {1 \over 2}\sqrt{1 + 4 \la^2}}~{\Gamma( 1 + 2 \alpha)~\Gamma(1 - \xi - 2 \beta)
\over \Gamma(1 - \xi + \alpha - \beta)~\Gamma(1 + \alpha - \beta)} \\ \nonumber
&+&
\big( r_{0} \big)^{- {1 \over 2} + {1 \over 2}\sqrt{1 + 4 \la^2}}~
\big( r \big)^{{1 \over 2} - {1 \over 2}\sqrt{1 + 4 \la^2}}~{\Gamma( 1 + 2 \alpha)~\Gamma(\xi - 1 + 2 \beta)
\over \Gamma(\xi + \alpha + \beta)~\Gamma(\alpha + \beta)}
\bigg].
\een
Because of the fact that {\it smooth matching} was provided,
the ratio of the integration constants may be written in the form
\be
\cB = {B_{+} \over B_{-}} = - {1 \over \pi~\big( r_{0} \big)^{\sqrt{1 + 4 \la^2}}~
\big( {1 \over 2} \tom \big)^{\sqrt{1 + 4 \la^2}}}
~{\Gamma(1 + {1 \over 2}{\sqrt{1 + 4 \la^2}})~\Gamma(1 - \xi - 2 \beta)
~\Gamma({1 \over 2}{\sqrt{1 + 4 \la^2}})~\Gamma(\xi + \alpha + \beta)~\Gamma(\alpha + \beta)
\over
\Gamma(1 - \xi + \alpha - \beta)~\Gamma(1 + \alpha - \beta)~\Gamma(\xi - 1 + 2 \beta)},
\label{bb}
\ee
while the absorption probability is given by the relation
\be
\mid \cA \mid^2 = {2i (\cB^{\ast} - \cB) \over \cB \cB^{\ast} + i (\cB^{\ast} - \cB) + 1}.
\label{abs}
\ee
The above relation can be simplified due to the fact that we are using the low-energy limit.
In this case $\cB \cB^{\ast} \gg i (\cB^{\ast} - \cB) \gg 1$.
Therefore, taking into account the dominant term in the denominator one arrives at the relation
\be
\mid \cA \mid^2 \simeq 2 i \bigg( {1 \over \cB} - {1 \over \cB^{\ast}} \bigg).
\ee
Using relation (\ref{bb}) it can be verified that the absorption probability implies the following:
\be
\mid \cA \mid^2 =
{8 \pi ~\om ~\tom^{\sqrt{1 + 4 \la^2}}~(r_{0})^{ \sqrt{1 + 4 \la^2} + 1}~ \Gamma(1 - (\xi + \beta))^2~
\Gamma(1 - \beta)^2 \over
2^{\sqrt{1 + 4 \la^2}}~(n + 1)~\sqrt{1 + 4 \la^2}~\Gamma(1 - 2 \beta - \xi)^2~
\Gamma({1 \over 2} \sqrt{1 + 4 \la^2})^2~(1 - \xi - 2 \beta)}.
\label{abso}
\ee
Eq.(\ref{abso}) is valid for the low-energy range of energy.
The low-energy approximation was used during matching the two asymptotic solutions
in the intermediate zone. However, the simplified analytical relation (\ref{abso})
is the result of series of Gamma function expansions appearing in Eq.(\ref{bb})
and therefore its validity is more restricted. In Table 1.
we presented the values of $\mid \cA \mid^2$ derived by using relation (\ref{abs})
and equation (\ref{abso}), as $\tom~r_{0}$ ranges from $0.01$ to $0.5$, for
$m=0.01,~B=1,~l=0$ and $n=1$. The results presented in Table 2. are valid for $n=2$, the other
calculation parameters are the same as in Table 1. 
One can conclude that,
the agreement between these two values in question is remarkable as $\tom~r_{0}$
reaches the value $0.5$, then the deviation between these values appears. 
\begin{table}[tn1]
\begin{center}
$\begin{array}{ccc}  \hline \hline
{\rule[-3mm]{0mm}{8mm}
\hspace*{0.5cm} \tom r_0 \hspace*{0.5cm}} & \hspace*{0.2cm}
| \cA |^2\,\,(\rm simplified\,\,expression~ Eq.(\ref{abso}))
\hspace*{0.2cm} & \hspace*{0.2cm} | \cA |^2\,\,(\rm given \,\, by ~ Eq.(\ref{abs}))
\hspace*{0.2cm} \\ \hline
{\rule[-2mm]{0mm}{6mm} 0.01} & 8.502 \times 10^{-9} & 8.502 \times 10^{-9} \\ 
{\rule[-2mm]{0mm}{6mm} 0.05} & 4.988 \times 10^{-6} & 4.988 \times 10^{-6}\\ 
{\rule[-2mm]{0mm}{6mm} 0.1} & 8.873 \times 10^{-5} & 8.873 \times 10^{-5} \\ 
{\rule[-2mm]{0mm}{6mm} 0.2} & 1.634 \times 10^{-3} & 1.632 \times 10^{-3} \\ 
{\rule[-2mm]{0mm}{6mm} 0.3} & 9.287 \times 10^{-3} & 9.244 \times 10^{-3} \\ 
{\rule[-2mm]{0mm}{6mm} 0.4} & 3.287 \times 10^{-2} & 3.233 \times 10^{-2} \\ 
{\rule[-2mm]{0mm}{6mm} 0.5} & 9.004 \times 10^{-2} & 8.612 \times 10^{-2}\\ \hline \hline
\end{array}$
\end{center}
\caption{Deviation between the values of the absorption probability given by the 
simplified and complete analytical expression, for $n=1$,~$m=0.01,~B=1,~l=0$ and different values of $\tom r_0 $.}
\end{table} 
\begin{table}[tn2]
\begin{center}
$\begin{array}{ccc}  \hline \hline
{\rule[-3mm]{0mm}{8mm}
\hspace*{0.5cm} \tom r_0 \hspace*{0.5cm}} & \hspace*{0.2cm}
| \cA |^2\,\,(\rm simplified\,\,expression~ Eq.(\ref{abso}))
\hspace*{0.2cm} & \hspace*{0.2cm} | \cA |^2\,\,(\rm given \,\, by ~ Eq.(\ref{abs}))
\hspace*{0.2cm} \\ \hline
{\rule[-2mm]{0mm}{6mm} 0.01} & 2.153 \times 10^{-11} & 2.153 \times 10^{-11} \\ 
{\rule[-2mm]{0mm}{6mm} 0.05} & 5.921 \times 10^{-8} & 5.921 \times 10^{-8}\\ 
{\rule[-2mm]{0mm}{6mm} 0.1} & 2.041 \times 10^{-6} & 2.041 \times 10^{-6} \\ 
{\rule[-2mm]{0mm}{6mm} 0.2} & 7.186 \times 10^{-5} & 7.186 \times 10^{-5} \\ 
{\rule[-2mm]{0mm}{6mm} 0.3} & 5.865 \times 10^{-4} & 5.863 \times 10^{-4} \\ 
{\rule[-2mm]{0mm}{6mm} 0.4} & 2.641 \times 10^{-3} & 2.638 \times 10^{-3} \\ 
{\rule[-2mm]{0mm}{6mm} 0.5} & 8.608 \times 10^{-3} & 8.571 \times 10^{-3}\\ \hline \hline
\end{array}$
\end{center}
\caption{Deviation between the values of the absorption probability given by the 
simplified and complete analytical expression, for $n=2$.
Other calculation parameters are the same as in Table 1.}
\end{table}

We remark that expression (\ref{abso}) is quite general, valid for both {\it bulk} and {\it brane} massive
Dirac fields. The case of {\it bulk} fermions is connected with the eigenvalues of Dirac operator
on a {\it transverse manifold} which is $S^{n + 2}$ sphere.
In case under investigation
one should consider
$(n + 2)$-dimensional sphere. The eigenvalues for spinor $\Psi$,
where found in Ref.\cite{cam96}. They imply the following relation:
\be
\la^2 = \bigg( l + { n + 2 \over 2} \bigg)^2,
\ee
where $l = 0, 1, \dots$
On the other hand, massive {\it brane} fermions, live on four-dimensional brane 
and propagate in the gravitational background provided by
\be
ds^2 = - f(r)dt^2 + {dr^2 \over f(r)} + r^2 d\Omega^2,
\ee
where $f = 1 - \bigg( r_{0} / r \bigg)^{n + 1}$, $r_{0}$ is the radius of the black hole event horizon.
This metric on the brane \cite{kan03a,kan03b}
arises when we project out all angular variables that parameterize the extra dimensions,
i.e., if we set $\phi_{i} = \pi/2$ for $i \ge 2$. 
In the above line element $n$ stands for the number of extra dimensions that can exist transverse to the considered brane.
Then, the eigenvalues on a {\it transverse} manifold are of the form $\la^2 = (l + 1)^2$.
\par
The other kind of $n$-dimensional black hole which can be considered by relation (\ref{abso})
is a tense brane black hole. It turned out that the
most examinations of extra-dimensions
black hole and their physics were devoted to the zero brane tension case. In principle
finite brane tension ought to modify the physics of a kind of black hole. 
The nonzero tension on the brane can curve the brane as well as the bulk.
It was shown that a
tense brane black hole is locally a higher-dimensional Schwarzschild solution \cite{kal06}
pierced by a tensional brane. This caused that a deficit angle appeared in the $(n + 2)$-dimensional
unit sphere line element.
\par
Some attempts to examine the problem in question were conducted. Namely,
studies of massless fermion excitation on a tensional three-brane were carried in Ref.\cite{cho08b},
while the late-time behaviour of massive scalar hair in the background of $n$-dimensional tense brane black hole was 
studied in \cite{rog08}.
Examinations of Hawking radiation of massive scalar fields in the background of a tense five and six-dimensional
black hole were carried in Ref.\cite{rog09}.
On the other hand,
emissions of massless scalar fields into the bulk from six-dimensional rotating black hole pierced by a three-brane
were studied in \cite{kob08}. Ref.\cite{dai07} was devoted to the numerical
studies of evaporation of massless scalar,
vector and graviton fields in the background of a six-dimensional tense brane black hole. 
Growing interests in codimensional-2 braneworlds lead us to modify
the gravitational action by implementing Gauss-Bonnet term or to consider black hole solutions
on a thin three-brane of codimension-2 \cite{cua08,cua08b}.
\par
In the case of a tense brane black hole $S^{n + 2}$ sphere is threaded by a codimension-2 brane, so
the range of the one of the angles, let us say, $\phi_{i}$ will be $ 0 \le \phi_{i} \le 2 \pi B$.
Parameter $B$ measures the deficit angle about axis parallel with the brane intersecting the sphere in question.
For such a kind of {\it transverse manifold} the eigenvalue for spinors will be of the form \cite{cho08}
\be
\la = k + {n + 2 \over 2} + \mid c \mid \bigg( {1 \over B} - 1 \bigg),
\ee
where $c = \pm 1/2,~\pm 3/2,~\dots$.

To complete this section let us turn to study the luminosity of the Hawking radiation for massive Dirac field.
It is provided by the expression
\ben \label{lum}
L &=& \int_{0}^{\infty} {d \om \over 2 \pi}~\mid \cA \mid^2_{l=0} {\om \over e^{\om \over T_{BH}} + 1} \\ \nonumber
&=& {4 ( r_{0} )^{\sqrt{ 1 + 4 \la^2} + 1}~\Gamma(1 - (\xi + \beta))^2~\Gamma(1 - \beta)^2 \over
2^{\sqrt{1 + 4 \la^2}}~(n + 1)~\sqrt{1 + 4 \la^2}~\Gamma(1 - 2 \beta - \xi)^2~
\Gamma({1 \over 2} \sqrt{1 + 4 \la^2})^2
~(1 - \xi - 2 \beta)} \\ \nonumber                    
&\times&
\bigg[ a_{1}~
(T_{BH})^{\sqrt{1 + 4 \la^2} + 3}~\zeta(\sqrt{1 + 4 \la^2} + 3)~\Gamma( \sqrt{1 + 4 \la^2} + 3) \\ \nonumber
&-& {a_{2}~m^2 \over 2}~\sqrt{1 + 4 \la^2}~
(T_{BH})^{\sqrt{1 + 4 \la^2} + 1}~\zeta(\sqrt{1 + 4 \la^2} + 1)~\Gamma(\sqrt{1 + 4 \la^2} + 1)
\bigg],
\een
where $T_{BH} = {(n+1) \over 4 \pi r_{0}}$ is the temperature of the black hole
while $\zeta$ is Riemann zeta function.  $a_{1}$ and $a_{2}$ are given by
\be
a_{1} = 1 - {1 \over 2^{2 + \sqrt{1 + 4 \la^2}}}, \qquad a_{2} =  1 - {1 \over 2^{ \sqrt{1 + 4 \la^2}}}.
\ee
In the above relation
(\ref{lum}) by $\la$ we mean $\la \mid_{l=0}$. In our low-energy approximation
the luminosity of the black hole Hawking radiation is calculated with respect to $l = 0$ mode. However,
for the sake of completeness we write in Eq.(\ref{lum}) the integral range from zero to infinity.
One should have in mind that our analysis has focused on the low-energy spectrum and the value of luminosity
is based on the lower part of the spectrum. It can happen that modifications may appear for high-energy part.

\section{The Absorption Probability and Hawking Radiation in the Spacetime of Higher Dimensional Black Hole}
In this section 
we present plots of the absorption probability and luminosity of Hawking radiation
for different parameters in different kinds of
$n$-dimensional black holes. 
The absorption probability is a dimensionless constant and it should range from 0 to 1 for the whole energy range.
Our considerations are confined to the low-energy limit therefore our plots will cover only a part of the whole
plot for $\mid \cA \mid^2$. The whole plot for $\mid \cA \mid^2$ can be found by numerical calculations
(see, e.g., Fig.1 in Ref.\cite{cho08} for the comparison of absorption probability obtained in the low-energy
limit and those get by various numerical approximation schemes). \\
In all our logarithmic plots we use {\it base-10 units}. One also elaborates the dependence of 
the absorption probability and luminosity of Hawking radiation for massive
{\it bulk} and {\it brane} fermions.

\subsection{Bulk Emission}
We begin our considerations with studies of massive {\it bulk} Dirac fermions emission.
In Fig.1 we plotted $\mid \cA \mid^2$ 
for massive fermion fields with respect to $\om$ for different number of extra dimensions $n=1, \dots 4$ 
in the background of $(n+4)$-dimensional Schwarzschild black holes and in the spacetime of tensional brane black hole.
We fixed the multipole number $l = 0$ for the Schwarzschild case, while 
for the tensional brane we put $k = 0,~c = 1/2$ and $B = 0.8$. 
Other calculation parameters are: $m=0.01$ and $r_0 = 1$. 
One should recall that the event horizon radius of a tense brane black hole
is connected with the Schwarzschild radius $r_{0}$ by the following relation:
\be
r_{TBBH} = {r_0 \over B^{1 \over n + 1}}.
\ee
It turned out that the absorption probability for massive {\it bulk} Dirac fermions for the considered tense brane black hole
is smaller comparing to the 
$(n + 4)$-dimensional Schwarzshild case.
\par
It was revealed in
Ref.\cite{cho08} using the WKBJ approximation and the Unruh method that the emission rate of 
the bulk massless fermions was dependent on the the dimensionality of the spacetime. The bigger is the spacetime dimension the 
smaller emission rate we get. It turned out that at a certain intermediate energy there was
a region where the emission rate became approximately independent of the dimension of underlying background.
Our Fig.1 confirms this tendency, i.e., for massive Dirac fermions the absorption probability decreases with the increase
of dimension of the spacetime.
\par
In Fig.2 we present the absorption probability versus $\om$ for $n = 5$, and 6, for different multipole numbers: 
$l = 0,~1$, and $2$, in the Schwarzschild background.
We also examine the case of a tense brane black hole and put
$k = 0,~1$, and~2. 
Other calculation parameters we take into account are: $m = 0.01$, $B = 0.8$, $r_0 = 1$, and $c = 1/2$. 
One can notice that the absorption probability decreases as the multipole number increases. 
\par
In Fig.3 we depict $\mid \cA \mid^2$ versus $\om$ 
for different mass of the Dirac massive fermion fields in
five and six-dimensional Schwarzschild black holes background. 
We consider the case when $m = 0,01,~0.1$, and~0.15. 
At the beginning  $\mid \cA \mid^2$
for massive Dirac fermions decreased as the mass of the field increased (the same situation was also revealed
in studies of the bulk absorption probability for scalars in the spacetime of Schwarzschild black hole 
located on a three-brane of finite tension (see Fig.4 in Ref.\cite{usa09})) but then there was a region
where the inverse situation took place. Contrary to the previuos behaviour the absorption probability
increased with the growth of the particles masses. This behaviour may be seen on the left panel
for five-dimensional spacetime and the same tendency occurs in six-dimensional case.
\par
The dependence of the absorption probability on the Schwarzschild black hole event horizon radius is presented in Fig.4.
We studied the case of $r_{0} = 1.2,~1$, and~0.8. One can find that the absorption probability increases
with the growth of radius of the black hole in question.\\
On the other hand,
inspection of Eq.(\ref{abso}) reveals that the absorption probability for massive {\it bulk} fermions
is strictly bounded with the parameter $B$ characterizing the brane tension. In Fig.5 we plotted $\mid \cA \mid^2$
versus $\om$ for different spacetime dimensionality $n = 5$, and ~6 and for different values of parameter $B$.
In the considerations we put $B = 1,~0.9,~0.8$, respectively. Fig.5 provides the fact that the smaller $B$ (brane tension increases)
the smaller absorption probability one gets. 
\par
In Ref.\cite{kob08} the emission bulk massless scalar fields from a six-dimensional rotating black 
hole pierced by a three-brane was numerically studied. For the low-energy region the authors find 
the dependence of absorption probability on parameter $B$. The curves presented in Fig.2 in \cite{kob08}
for the case of angular parameter equal to zero has the same tendency as our Fig.5 (right panel). 
The same situation takes place 
when one considers massive scalar fields emitted by a tense brane black hole \cite{rog09}.
So one can conclude that the brane tension suppresses the emission of massless scalars, massive scalars as well as
massive bulk fermions.
\par
In Fig.6 
we present the $\mid \cA \mid^2$ for five and six-dimensional tensional brane black holes 
for different $c=1/2,~3/2,~5/2$, respectively. Other calculation parameters are: $m=0.01$, $r_0=1$, $k=0$ and $B=0.9$. 
It occurs that the 
bigger $c$ is the smaller absorption probability one obtains.

\subsection{Emission on the Brane}
Now, we proceed to study
the absorption probability of Dirac massive fermions propagating on the brane. 
In the left panel of Fig.7 we present the brane absorption probability for fixed $l=0$ for 
different spacetime dimensionalities $n=1, \dots 4$. 
It turns out that the absorption probability increases with the increase of spacetime dimensionality $n$
for the brane black hole which all angular 
variables parameterizing the extra dimensions are projected out. The same behaviour of $\mid \cA \mid^2$
was shown for massless {\it brane} fermions \cite{kan03b} in the spacetime
of rotating brane black hole. These results were also confirmed numerically in Ref.\cite{cas07} where
it was also revealed that any increase of spacetime dimensionality or angular parameter of black hole
enhances the massless fermion emission rate of the considered black hole.
Studies of radial Teukolsky Eq. conducted in Ref.\cite{ida06} also confirmed the tendency of behaviour we obtained
(see Fig.1 from Ref.\cite{ida06} for the rotation parameter equal zero depicting greybody factor for brane localized fermions).
\par
In the right panel of Fig.7 we plotted the $\mid \cA \mid^2$ as a function of $\om$ 
for different multipole number $l=0,~1$, and~2 in five and six-dimensional spacetimes. 
It happens that the brane absorption probability decreases as the multipole number increases, as was the case for the bulk radiation.
\par
Next, in Fig.8
we examined 
the behaviour of the absorption probability on the brane for different masses of the fermion particles.
We put $m = 0.01,~0.1,~0.15$, respectively. 
At the beginning $\mid \cA \mid^2$ decreases as the mass of fermions increases but there is a region where
the situation changes, i.e., for the increase of mass one gets also increase of the absorption probability.
The same situation was observed in the bulk case.
\par
On the other hand, in Fig.9 we plotted the dependence of $\mid \cA \mid^2$
on the radii of black hole event horizons. One studies the cases of $r_{0} = 1.2,~1$, and~0.8. It can be concluded 
that the absorption probability for massive {\it brane} Dirac fermions
increases as the event horizon radius increases.
The same behaviour we have in the case of massive {\it bulk} fermions.

\subsection{Luminosity} 
Finally, we plotted the luminosity of the Hawking radiation $L$ as a function of $m$ 
for different number of extra spacetime dimensions $n=1, \dots 4$. 
We plot the luminosity of the Hawking radiation for the mode $l = 0$ which plays the dominant role in a 
greybody factor in the low-energy approximation.
In the left panel of Fig.10 
we present the bulk absorption probability for Schwarzschild and tensional brane black holes. 
In the right one, we plotted the brane absorption probability in the background of higher dimensional 
Schwarzschild black holes. 
One can remark that the luminosity of Hawking radiation increases with the increase of $n$, spacetime dimensionality for
massive {\it bulk} and {\it brane} fermion emission. The other salient feature is that luminosity for {\it brane}
massive fermion is substantially higher comparing to the luminosity for {\it bulk} fermions for the corresponding 
dimensionality of spacetime.
Of course this behaviour is valid only for the low-energy limit. The behaviour of the luminosity valid for the whole
energy range should be obtained by numerical studies. However our analytical results tell us about the tendency of behaviour
of the luminosity in this energy limit. In Ref.\cite{cho08} the massless fermion emission rate
was studied numerically and it was shown that for the brane localized emission one obtained the same tendency of 
behaviour as ours, i.e., Hawking radiation increased with the growth of the dimensionality of the spacetime.
It was also found that in the intermediate energy range the emission rate became independent on the dimensionality.
However this case is beyond our approximation scheme. 
\par 
The same tendency of behaviour was revealed in Ref.\cite{cas08} for the massless scalar 
emission rate in the spacetime of $n$-dimensional rotating black hole. Namely, the authors find that 
for the rotational parameter put to zero the total emission depends strongly on the number 
of dimensions both for bulk and brane emission. It turned out that on average, more energetic 
emission is in the bulk than on the brane. Unfortunately, our plots of the luminosity 
for massive Dirac fermions can not confirm this behaviour 
valid for massless scalars because of the considered energy limit. \\
The dependence of a flux emission spectra of massless fermion on the brane emitted from rotating 
black hole on the spacetime dimension was also confirmed in Ref.\cite{cas07}.
\par
Thus, summing it all up we can conclude that in the considered energy limit our key result is that the luminosity for massive
Dirac fermions strongly depends on the bulk and brane dimensions. But as one can try to consider the whole energy range these
results should be taken with a large grain of salt.

\section{Conclusions}
In this paper we studied Hawking emission of massive Dirac fermion fields in the
spacetime of static $(4 + n)$-dimensional black holes. 
We elaborated the case of $(4 + n)$ Schwarzschild black hole, $(4 + n)$ tense brane black hole
as well as propagation of {\it brane} fermions in the spacetime of brane black hole when all 
angular variables parameterizing the extra dimensions
in the considered line element were projected out. It happened that the treatment of 
Dirac fermions in spherically symmetric spacetime was simplified to great extent due to the few 
properties of the Dirac operator. The crucial quantities characterizing emission of massive Dirac fields
are dependent on the eigenvalues $\la$ of the Dirac operator on the so-called {\it transverse manifolds}.
We have elaborated analytically
equations of motion for the aforementioned degrees of freedom and by means of {\it matching technique}
one finds in the low-energy limit an analytical expression for the absorption cross section, luminosity of Hawking radiation.
We derive quite general formulae for these quantities which enable one to treat all these spacetimes in question.
Our analytical considerations are supplemented by plots
of $\mid \cA \mid^2$ and the luminosity of Hawking radiation $L$.
We found that  $\mid \cA \mid^2$ for {\it bulk} massive Dirac fermions in the spacetime of $(n + 4)$-dimensional
Schwarzschild black hole and in the background of a tense brane black hole decreases with increasing of the 
dimensionality of the considered spacetime. In higher dimensional Schwarzschild spacetime the absorption probability decreases as the 
multiple number $l$ increases. For this spacetime it was also revealed that  $\mid \cA \mid^2$ depended 
on mass of the emitted field.
First, at the beginning of the energy range, we have the situation that the bigger mass of Dirac fermion is
the smaller $\mid \cA \mid^2$ one gets. Then, this tendency changes and we observe the increase of the
absorption probability with the growth of mass of particle in question.
It was also observed that the absorption probability increased with the increase of the radius of the event horizon.
We note that $\mid \cA \mid^2$ reveals its dependence on parameter characterizing tense brane black hole. Namely,
the smaller $B$ one considers the the smaller $\mid \cA \mid^2$ one gets. One ought to have in mind that $B$ is bound 
with the brane tension, i.e., the smaller $B$ the greater tension is exerted on the brane black hole. In the case of
a tense brane black hole the growth of a multipole number $c$ implies the decreasing of the absorption probability.
\par
We also studied the case of {\it brane} massive Dirac fermion field. It was shown that $\mid \cA \mid^2$
increased with the increase of $n$ for brane black hole which all angular variables parameterizing the extra 
dimensions were projected out. The same character of behaviour was shown in Ref.\cite{kan03b} for the case of massless
{\it brane} fermions. It was also noticed that $\mid \cA \mid^2$ for different masses of {\it brane} fermions 
behaved in the similar way as in the case of {\it bulk} massive
Dirac fermions.
On the other hand, $\mid \cA \mid^2$ analyzed for different event horizon radius also reacts in the similar
way as in the case of {\it bulk} fermions.
Finally, analyzing luminosity of Hawking radiation for massive {\it brane} Dirac fermions it was found that the {\it brane}
luminosity was substantially higher comparing to the luminosity of Hawking radiation for {\it bulk} massive fermions for the
corresponding values of spacetime dimensionality $n$.
One should remark that the behaviour of the luminosity is valid for the low-energy limit, when $\tom~r_{0} \ll 1$. 

 

\begin{acknowledgments}
This work was partially financed by the Polish budget funds in 2009 year as
the research project.
\end{acknowledgments}


\pagebreak

\begin{figure}
\begin{center}
  \includegraphics[width=0.85\textwidth]{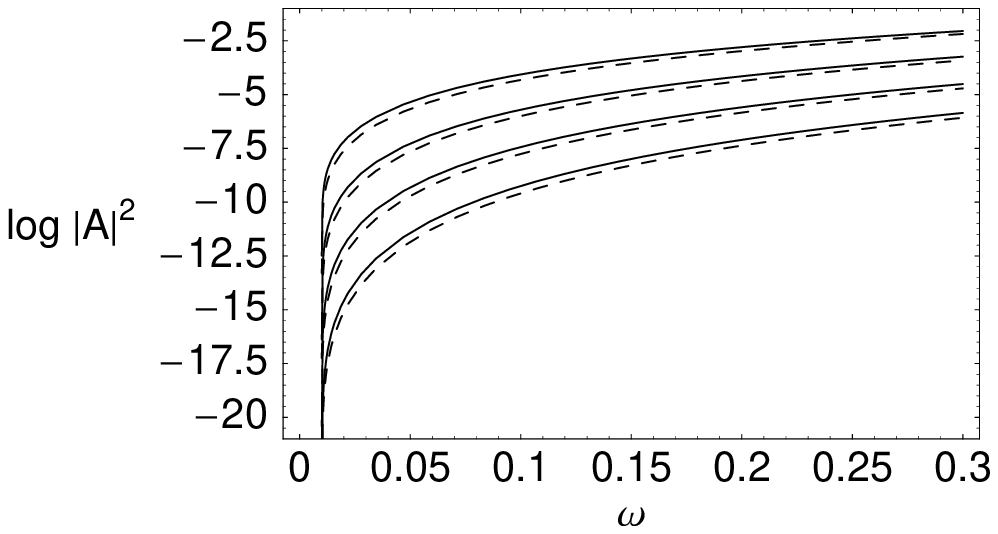}
\end{center}
\caption{Absorption probability $\mid \cA \mid^2$ for bulk Dirac fermions with mass $m$ as a function of
$\om$, for different space dimensionality $n = 1,~2,~3,~4$ (curves from the top to the bottom, respectively). 
The solid lines represent the case of $(n+4)$-dimensional Schwarzschild black holes, the dashed lines 
represent $(n + 4)$-dimensional tense brane black holes with $B = 0.8$.
Other calculation parameters are:~$c =1/2,~k = 0,~l = 0,~r_0 = 1$, and $m = 0.01$.}
\label{fig:fig1}
\end{figure}

\pagebreak
\pagebreak

\begin{figure}
\begin{center}
  \includegraphics[width=0.85\textwidth]{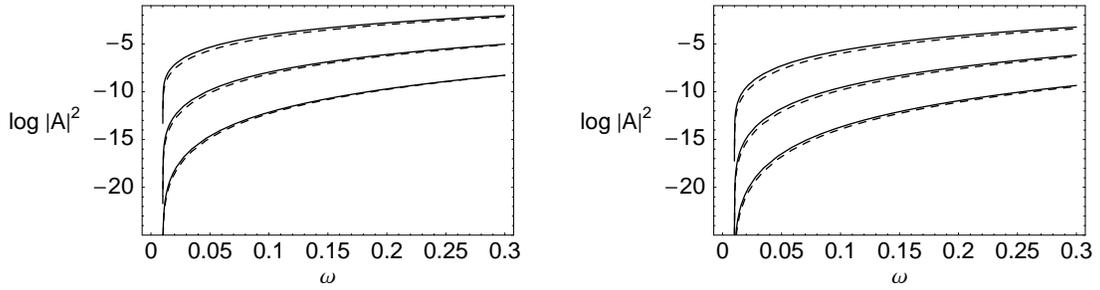}
\end{center}
\caption{Absorption probability $\mid \cA \mid^2$ for massive Dirac field as a function of
$\om$, for different values of $l = 0,~1,~2$ (solid lines, curves from the top to the bottom, respectively) 
for Schwarzschild black holes 
and for $k=0,~1,~ 2$ for tense brane black holes (dashed lines, curves from the top to the bottom, respectively), 
for five and six-dimensional cases (left and right panel, respectively).
The rest of the calculation parameters are the same as in Fig.1.}
\label{fig:fig2}
\end{figure}

\pagebreak

\begin{figure}
\begin{center}
  \includegraphics[width=0.85\textwidth]{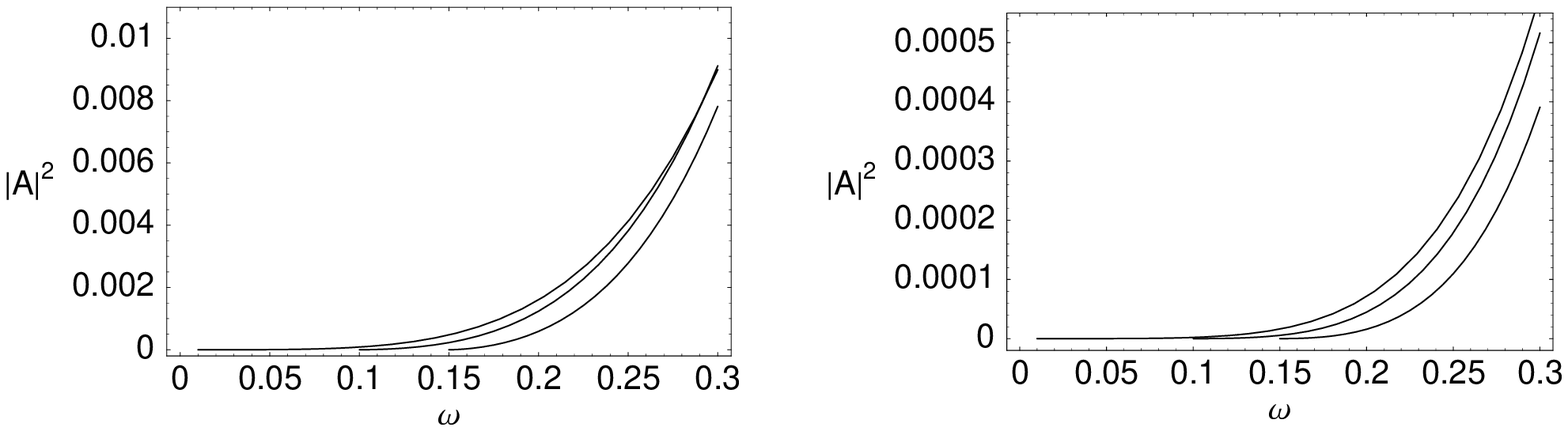}
\end{center}
\caption{Absorption probability $\mid \cA \mid^2$ for massive fermion
particles in a spacetime of five and six-dimensional Schwarzshild black hole (left and right panel, respectively),
for different values of the mass of Dirac fermions $m = 0.01,~0.1,~ 0.15$ (curves from the top to the bottom, respectively).
Other calculation parameters are:~$l = 0$, and $r_0 = 1$.}
\label{fig:fig3}
\end{figure}

\pagebreak

\begin{figure}
\begin{center}
  \includegraphics[width=0.85\textwidth]{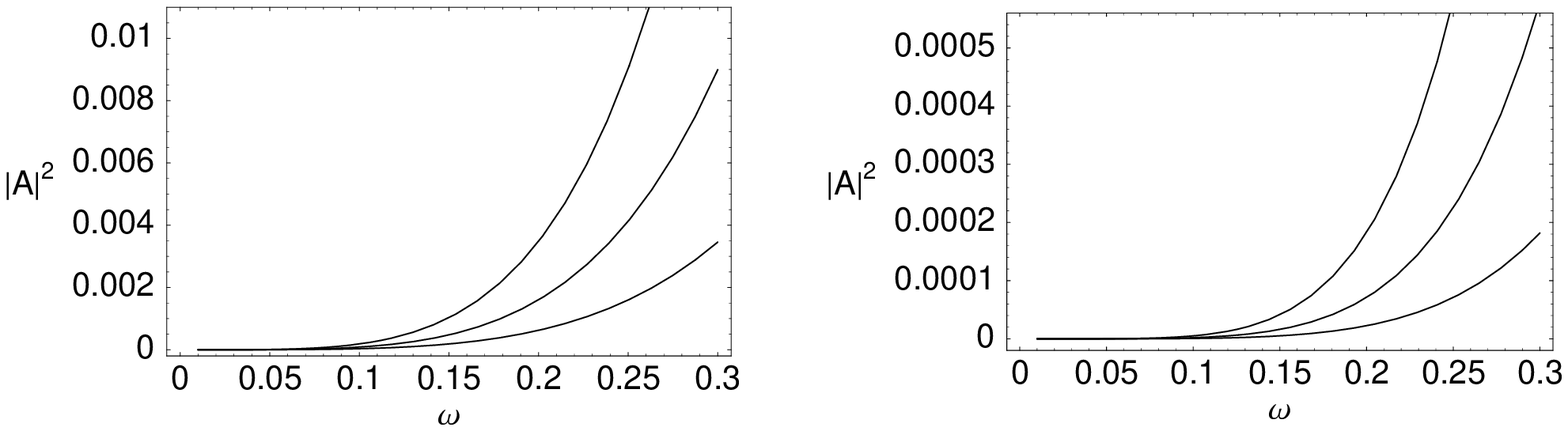}
\end{center}
\caption{Absorption probability $\mid \cA \mid^2$ for massive fermion
particles in a spacetime of five and six-dimensional Schwarzshild black hole (left and right panel, respectively).
Curves from the top to the bottom are for different values of black hole event horizon radius
$r_0 = 1.2,~1$, and 0.8.
Calculation parameters are:~$l = 0$, and $m=0.01$.}
\label{fig:fig4}
\end{figure}
\pagebreak

\begin{figure}
\begin{center}
  \includegraphics[width=0.85\textwidth]{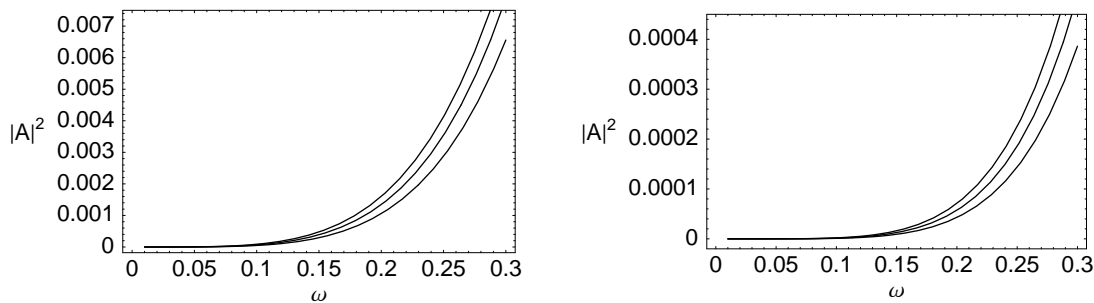}
\end{center}
\caption{Absorption probability $\mid \cA \mid^2$ versus $\om$ for 
different values of $B = 1,~0.9,~0.8$ (curves from the top to the bottom, respectively) and
for different spacetime dimensionality $n = 1$, and 2 (panels from left to the right) for tense brane black holes.
The rest of the parameters are the same as in Fig.1}
\label{fig:fig5}
\end{figure}

\pagebreak

\begin{figure}
\begin{center}
  \includegraphics[width=0.85\textwidth]{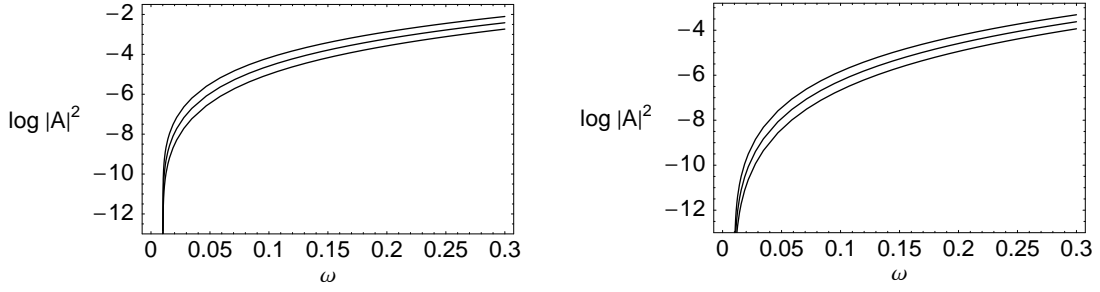}
\end{center}
\caption{Absorption probability $\mid \cA \mid^2$ versus $\om$ for 
massive fermion field for different values of $c = 1/2,~3/2$, and $5/2$ (curves from the top to the bottom, respectively) 
and
for different spacetime dimensionalities $n = 1,~2$ (panels from left to the right) in the background of a tense brane black hole.
Other calculation parameters are: $m=0.01,~r_0 = 1,~k=0$, and $B=0.9$.}
\label{fig:fig6}
\end{figure}

\pagebreak

\begin{figure}
\begin{center}
  \includegraphics[width=0.85\textwidth]{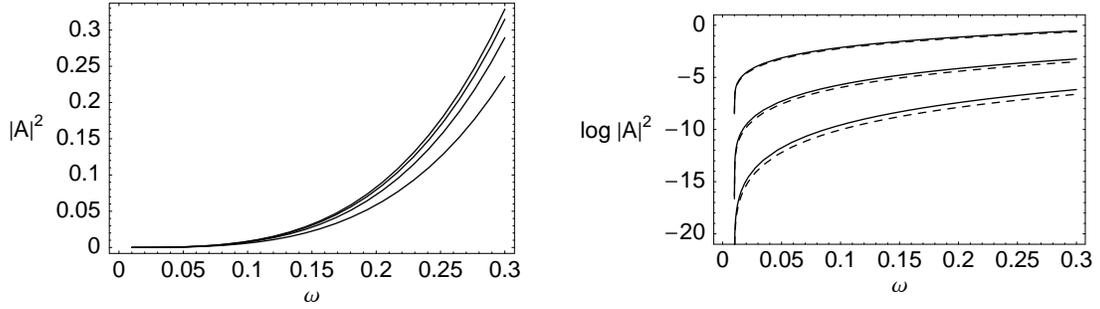}
\end{center}
\caption{Left panel: absorption probability $\mid \cA \mid^2$ versus $\om$ for 
massive brane fermions for fixed $l=0$ and different spacetime dimensionalities $n=1,~2,~3,~4$ 
(curves from the bottom to the top, respectively). Right panel: absorption probability 
for different multipole numbers  $l=0,~1,~2$ (curves from the top to the bottom, respectively) for five and 
six-dimensional cases (dashed and solid lines, respectively). Other calculation parameters: $m=0.01$, and $r_0=1$.}
\label{fig:fig7}
\end{figure}

\pagebreak

\begin{figure}
\begin{center}
  \includegraphics[width=0.85\textwidth]{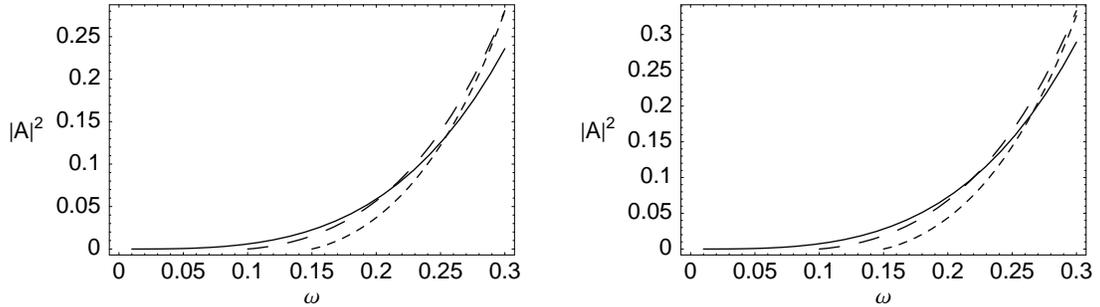}
\end{center}
\caption{Absorption probability $\mid \cA \mid^2$ for massive brane fermion
particles in a spacetime of five and six-dimensional Schwarzschild black holes (left and right panel, respectively),
for different values of $m = 0.01,~0.1,~0.15$ (curves from the top to the bottom, respectively).
Other calculation parameters are:~$l = 0$, and $r_0 = 1$.}
\label{fig:fig8}
\end{figure}

\pagebreak

\begin{figure}
\begin{center}
  \includegraphics[width=0.85\textwidth]{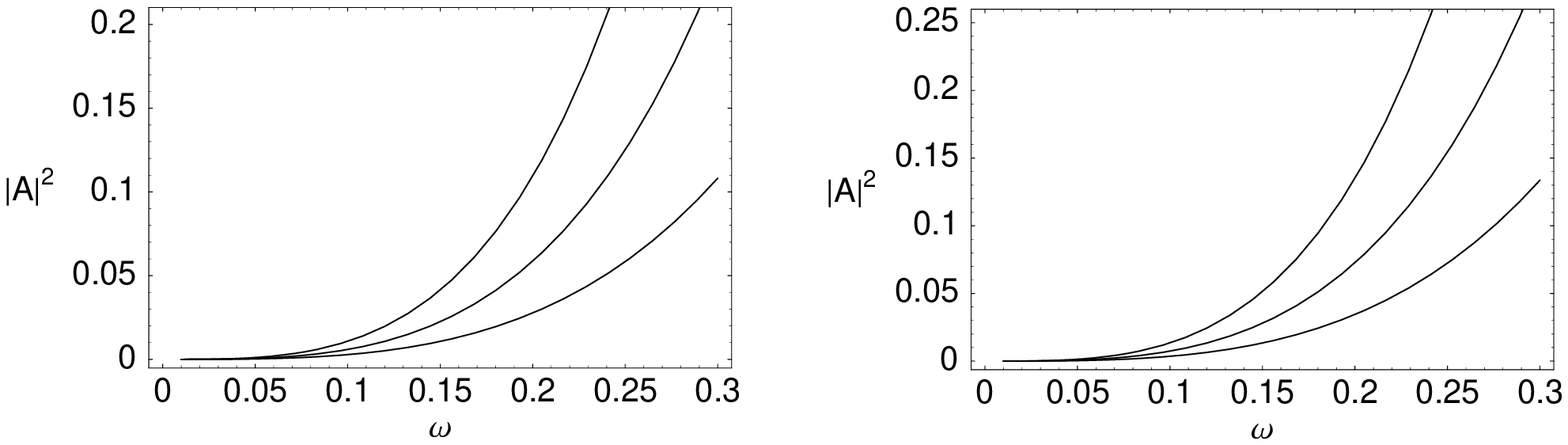}
\end{center}

\caption{Absorption probability $\mid \cA \mid^2$ for massive brane fermion
particles in a spacetime of five and six-dimensional Schwarzschild black holes (left and right panel, respectively).
Curves from the top to the bottom are for different values of black hole event horizon radius
$r_0 = 1.2,~1$, and $0.8$.
Other calculation parameters are:~$l = 0$, and $m=0.01$.}
\label{fig:fig9}
\end{figure}


\pagebreak

\begin{figure}
\begin{center}
  \includegraphics[width=0.85\textwidth]{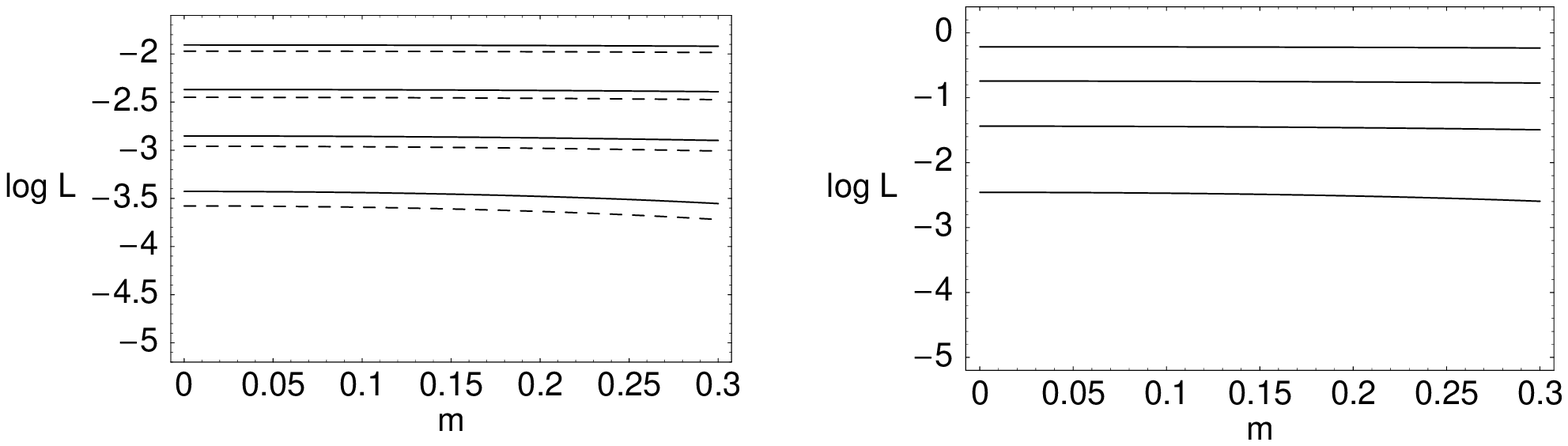}
\end{center}
\caption{The luminosity of the Hawking fermion radiation $L$ versus $m$ propagating in the bulk (left panel) 
on a Schwarzschild black hole spacetime (solid lines) and on a tense brane black hole spacetime with $B=0.9$
and $k=0,~c=1/2$~ (dashed lines)
for different space dimensionalities $n = 1,~2,~3,~4$ (curves from the bottom to the top, respectively). 
Right panel: the luminosity of Hawking radiation  $L$ versus $m$ propagating 
on the brane for different space dimensionalities $n = 1,~2,~3,~4$ (curves from the bottom to the top, respectively).
For all cases we put $r_0=1$. }
\label{fig:fig10}
\end{figure}


\begin{thebibliography}{99}
%
\def\cmp#1#2#3{{ Commun. Math. Phys.} {\bf #1}, #2 (#3)}
\def\lmp#1#2#3{{ Lett. Math. Phys.} {\bf #1}, #2 (#3)}
\def\hpa#1#2#3{{ Hell. Phys. Acta} {\bf #1}, #2 (#3)}
\def\grg#1#2#3{{ Gen. Rel. Grav.} {\bf #1}, #2 (#3)}
\def\pr#1#2#3{{ Phys. Rev.} {\bf #1}, #2 (#3)}
\def\prl#1#2#3{{ Phys. Rev. Lett.} {\bf #1}, #2 (#3)}
\def\prd#1#2#3{{ Phys. Rev. D} {\bf #1}, #2 (#3)}
\def\pl#1#2#3{{ Phys. Lett} {\bf #1}, #2 (#3)}
\def\pla#1#2#3{{ Phys. Lett. A} {\bf #1}, #2 (#3)}
\def\plb#1#2#3{{ Phys. Lett. B} {\bf #1}, #2 (#3)}
\def\prep#1#2#3{{ Phys. Reports} {\bf #1}, #2 (#3)}
\def\phys#1#2#3{{ Physica} {\bf #1}, #2 (#3)}
\def\jcp#1#2#3{{ J. Comput. Phys.} {\bf #1}, #2 (#3)}
\def\jmp#1#2#3{{ J. Math. Phys.} {\bf #1}, #2 (#3)}
\def\jpm#1#2#3{{ J. Phys. A: Math. Gen.} {\bf #1}, #2 (#3)}
\def\cpr#1#2#3{{ Computer Phys. Rept.} {\bf #1}, #2 (#3)}
\def\cqg#1#2#3{{ Class. Quantum Grav.} {\bf #1}, #2 (#3)}
\def\cma#1#2#3{{ Computers Math. Applic.} {\bf #1}, #2 (#3)}
\def\mc#1#2#3{{ Math. Compt.} {\bf #1}, #2 (#3)}
\def\apj#1#2#3{{ Astrophys. J.} {\bf #1}, #2 (#3)}
\def\apjs#1#2#3{{ Astrophys. J. Suppl.} {\bf #1}, #2 (#3)}
\def\acta#1#2#3{{ Acta Astronomica} {\bf #1}, #2 (#3)}
\def\jgp#1#2#3{{ J.Geom.Phys.} {\bf #1}, #2 (#3)}
\def\apl#1#2#3{{Ann. Physik. (Leipzig)} {\bf #1}, #2 (#3)}
\def\anp#1#2#3{{Ann. Phys. } {\bf #1}, #2 (#3)}
\def\sa#1#2#3{{ Sov. Astro.} {\bf #1}, #2 (#3)}
\def\sia#1#2#3{{ SIAM J. Sci. Statist. Comput.} {\bf #1}, #2 (#3)}
\def\aa#1#2#3{{ Astron. Astrophys.} {\bf #1}, #2 (#3)}
\def\mnras#1#2#3{{ Mon. Not. R. astr. Soc.} {\bf #1}, #2 (#3)}
\def\npb#1#2#3{{ Nucl. Phys. B} {\bf #1}, #2 (#3)}
\def\prsla#1#2#3{{ Proc. R. Soc. London, Ser. A} {\bf #1}, #2 (#3)}
\def\jhep#1#2#3{{ JHEP} {\bf #1}, #2 (#3)}
\def\nuc#1#2#3{{Nuovo Cimento B } {\bf #1}, #2 (#3)}
\def\ijmpd#1#2#3{{Int. J. Mod. Phys. D} {\bf #1}, #2 (#3)}
\def\ijmpa#1#2#3{{Int. J. Mod. Phys. A} {\bf #1}, #2 (#3)}
\def\atmp#1#2#3{{Adv. Theor. Math. Phys.} {\bf #1}, #2 (#3)}
\def\ptps#1#2#3{{Prog. Theor. Phys. Suppl.} {\bf #1}, #2 (#3)}
\def\lmp#1#2#3{{Lett. Math. Phys. } {\bf #1}, #2 (#3)}
\def\epjc#1#2#3{{Eur. Phys. J. C} {\bf #1}, #2 (#3)}
\def\hepph#1#2{{ hep-ph }{\bf #1} (#2)}
\def\hepth#1#2{{ hep-th }{\bf #1} (#2)}
\def\grqc#1#2{{ gr-qc }{\bf #1} (#2)}
\def\ibid#1#2#3{{ {\it ibid.} }{\bf #1}, #2 (#3)}
%
\bibitem{ark98}
N.Arkani-Hamed, S.Dimopoulos, and G.R.Dvali, \plb{429}{263}{1998},\\
I.Antoniadis, N.Arkani-Hamed, S.Dimopoulos, and G.R.Dvali, \ibid{436}{257}{1998}.
\bibitem{ran99}
L.Randall and R.Sundrum, \prl{83}{3370}{1999},\\
L.Randall and R.Sundrum, \ibid{83}{4690}{1999}.


\bibitem{kan03a}
P.Kanti and J.March-Russell, \prd{66}{024023}{2003}.
\bibitem{kan03b}
P.Kanti and J.March-Russell, \prd{67}{104019}{2003}.
\bibitem{par06}
D.K.Park, \plb{633}{613}{2006}.
\bibitem{har03}
C.M.Harris and P.Kanti, \jhep{10}{014}{2003},\\
P.Kanti, \ijmpa{19}{4899}{2004},\\
P.Argypes, S.Dimopoulos, and J.March-Russell, \prb{441}{96}{1998}.
\bibitem{jun05}
E.Jung and D.K.Park, \npb{717}{272}{2005},\\
E.Jung and D.K.Park, \cqg{21}{3717}{2004},\\
E.Jung, S.H.Kim, and D.K.Park, \plb{586}{390}{2004},\\
E.Jung, S.H.Kim, and D.K.Park, \ibid{614}{78}{2005},\\
E.Jung, S.H.Kim, and D.K.Park, \ibid{615}{273}{2005},\\
E.Jung, S.H.Kim, and D.K.Park, \ibid{619}{347}{2005},\\
G.Duffy, C.Harris, P.Kanti, and E.Winstanley, \jhep{02}{049}{2005},\\
M.Casals, P.Kanti, and E.Winstanley, \ibid{02}{051}{2006}.
\bibitem{car06}
V.Cardoso, M.Cavaglia, and L.Gualtieri, \prl{96}{071301}{2006},\\
V.Cardoso, M.Cavaglia, and L.Gualtieri, \jhep{02}{021}{2006}.

\bibitem{cre06}
S.Creek, O.Efthimiou, P.Kanti, and K.Tamvakis, \plb{635}{39}{2006}.
\bibitem{cre07}
D.Ida, K.Oda, and S.C.Park, \prd{67}{064025}{2003},\\
D.Ida, K.Oda, and S.C.Park, \ibid{71}{124039}{2005},\\
S.Creek, O.Efthimiou, P.Kanti, and K.Tamvakis, \prd{75}{084043}{2007},\\
S.Creek, O.Efthimiou, P.Kanti, and K.Tamvakis, \ibid{76}{104013}{2007},\\
S.Creek, O.Efthimiou, P.Kanti, and K.Tamvakis, \plb{656}{102}{2007},\\
E.Jung and D.K.Park, \npb{731}{171}{2005}.


\bibitem{gra08}
J.Grain and A.Barrau, \epjc{53}{641}{2008}.




\bibitem{fro03}
V.Frolov and D.Stojkovic, \prd{67}{084004}{2003}.

\bibitem{cas08}
M.Casals, S.Dolan, P.Kanti, and E.Winstanley, \jhep{06}{071}{2008}.
\bibitem{cas07}
M.Casals, S.Dolan, P.Kanti, and E.Winstanley, \jhep{03}{019}{2007}.
\bibitem{ida06}
D.Ida, K.Oda, S.C.Park, \prd{73}{124022}{2006}.
\bibitem{cho08}
H.T.Cho, A.S.Cornell, J.Doukas, W.Naylor, \prd{77}{016004}{2008}.



\bibitem{usa09}
U.al-Binni and G.Siopsis, {\it Particle Emission from a Black Hole on a Tense Codimension-2 Brane},
\hepth{0902.2194}{2009}.


\bibitem{che08}
S.Chen, B.Wang, and J.Jing, \prd{78}{064030}{2008}.
\bibitem{che08b}
S.Chen, B.Wang, and R.Su, \prd{77}{024039}{2008}.
\bibitem{wu08}
S.F.Wu, S.Yin, G.H.Yang, P.M.Zhang, \prd{78}{084010}{2008}.
\bibitem{che08c}
S.Chen, B.Wang, and R.Su, \prd{77}{124011}{2008}.
\bibitem{cve98}
M.Cvetic and F.Larsen, \prd{57}{6297}{1998}.
\bibitem{das97}
S.R.Das, G.W.Gibbons, and S.D.Mathur, \prl{78}{417}{1997}.
\bibitem{jun04}
E.Jung, S.H.Kim, and D.K.Park, \jhep{05}{0409}{2004}.

\bibitem{gib08a}
G.W.Gibbons and M.Rogatko, \prd{77}{044034}{2008}.
\bibitem{gib08b}
G.W.Gibbons, M.Rogatko, and A.Szyplowska, \prd{77}{064024}{2008}.
\bibitem{mod08}
R.Moderski and M.Rogatko, \prd{77}{124007}{2008}.
\bibitem{zum62}
B.Zumino, \jmp{3}{1055}{1962}.

\bibitem{abr66}
{\it Handbook of Mathematical Functions}, edited by M.Abramowitz and I.Stegun (Academic, New York, 1966).


\bibitem{cam96}
R.Camporesi and A.Higuchi, \jgp{20}{1}{1996}.
\bibitem{kal06}
N.Kaloper and D.Kiley, \jhep{03}{077}{2006}.
\bibitem{cho08b}
H.T.Cho, A.S.Cornell, J.Doukas, and W.Naylor, \prd{77}{041502(R)}{20008}.
\bibitem{rog08}
M.Rogatko and A.Szyplowska, {\it Evolution of Massive Scalar Fields in the Spacetime of a Tense 
Brane Black Hole}, \hepth{0812.1644}{2008}.
\bibitem{rog09}
M.Rogatko and A.Szyplowska, research in progress.

\bibitem{kob08}
T.Kobayashi, M.Nozawa, and Y.Takamizu, \prd{77}{044022}{2008}.
\bibitem{dai07}
D.C.Dai, N.Kaloper, G.D.Starkman, and D.Stojkovic, \prd{75}{024043}{2007}.
\bibitem{cua08}
B.Cuadros-Melgar, E.Papantonopoulos, M.Tsoukalas, and V.Zamarias, \prl{100}{221601}{2008}.
\bibitem{cua08b}
B.Cuadros-Melgar, E.Papantonopoulos, M.Tsoukalas, and V.Zamarias, {\it Black Holes on Thin 3-branes 
of Codimension 2 and Their Extension into the Bulk}, \hepth{0804.4459}{2008}.









                           
\end{thebibliography}
\end{document}